\documentclass[twoside]{article}

\usepackage{tabulary,graphicx,times,caption,fancyhdr,amsfonts,amssymb,amsbsy,latexsym,amsmath}
\usepackage[utf8]{inputenc}
\pdfoutput=1
\usepackage{url,multirow,morefloats,floatflt,cancel,tfrupee,textcomp,colortbl,xcolor,pifont}
\usepackage[nointegrals]{wasysym}
\usepackage{graphics}
\usepackage{ulem}
\usepackage{multirow}
\usepackage{float}
\usepackage{inputenc}
\usepackage{xspace}
\usepackage{url}
\usepackage{epstopdf}
\usepackage{amsfonts}
\usepackage{amsmath}
\usepackage{amssymb}
\usepackage{extarrows}
\graphicspath{ {./images/} }
\allowdisplaybreaks

\makeatletter

\usepackage{ifxetex}
\ifxetex\else
  \usepackage{dblfloatfix}
\fi

\@ifundefined{subparagraph}{
\def\subparagraph{\@startsection{paragraph}{5}{2\parindent}{0ex plus 0.1ex minus 0.1ex}%
{0ex}{\normalfont\small\itshape}}%
}{}

\def\URL#1#2{\@ifundefined{href}{#2}{\href{#1}{#2}}}

\def\UrlOrds{\do\*\do\-\do\~\do\'\do\"\do\-}%
\g@addto@macro{\UrlBreaks}{\UrlOrds}

\makeatother

\usepackage[paperheight=11in,paperwidth=8.3in,margin=2.5cm,headsep=.7cm,top=2.5cm]{geometry}
\usepackage[T1]{fontenc}

\renewenvironment{abstract}
	{\trivlist\item[]\leftskip0pt\par\vskip4pt\noindent
  	\textbf{\abstractname}\mbox{\null}\\}
	{\par\noindent\endtrivlist}

\def\keywords#1{\par\medskip\par\noindent\textbf{Keywords}: #1\par}

 \date{} \emergencystretch 8pt

\makeatletter
\def\author#1{\gdef\@author{\hskip-\tabcolsep%
	\parbox{\textwidth}{\raggedright\bfseries#1\\[1pc]}}}
\def\address[#1]#2{\g@addto@macro\@author{\\\hskip-\tabcolsep\parbox{\textwidth}{\raggedright%
	\normalsize\normalfont\textsuperscript{#1}#2}}}
\let\addresslink\textsuperscript
\def\correspondence#1{\g@addto@macro\@author{\\\hskip-\tabcolsep\parbox{\textwidth}{\raggedright%
	\vspace*{10pt}\normalsize\normalfont~\\#1~\\[12pt]}}}
\def\email#1{\g@addto@macro\@author{\\\hskip-\tabcolsep\parbox{\textwidth}{\raggedright%
	\normalsize\normalfont Emails: #1}}}

\def\title#1{\gdef\@title{\vspace*{-30pt}%
	\raggedright\textbf{\@journaltitle}~\\%
  \raggedright\bfseries\ifx\@articleType\@empty\vspace*{20pt}\else%
  \vspace*{20pt}\@articleType\vspace*{20pt}\\\fi#1}}
\let\@journaltitle\@empty \def\journaltitle#1{\gdef\@journaltitle{{\normalfont\itshape#1}}}
\let\@articleType\@empty \def\articletype#1{\gdef\@articleType{{\normalfont\itshape#1}}}

\let\@runningHead\@empty \def\RunningHead#1{\gdef\@runningHead{{\normalfont #1}}}

\usepackage{fancyhdr}
\fancypagestyle{headings}{\fancyhf{}
  \fancyhead[R]{\itshape\@runningHead}
  \fancyfoot[C]{\thepage}}
\fancypagestyle{plain}{%
	\fancyhf{}\fancyhead[R]{Prepared for submission to International Journals of High Energy Physics}
  \fancyfoot[C]{\thepage}}
\makeatother

\usepackage[numbers]{natbib}

\usepackage{float,xcolor}

\journaltitle{International Journals of High Energy Physics}
\articletype{Research Article} % Research Article/Review Article/Clinical Study

\begin{document}

\title{Fermions coupled to Chern-Simons gauge field or imaginary chemical potential and the Bloch theorem}

\author{
		Evangelos G. Filothodoros\addresslink{1}}		
% Affiliation
\address[1]{Institute of Theoretical Physics, Aristotle University of Thessaloniki, Thessaloniki, Greece.}

\correspondence{Correspondence should be addressed to 
    	Evangelos G. Filothodoros; efilotho@physics.auth.gr,vagfil79@gmail.com}

% Emails of author

% Running Head

\maketitle 

% Abstract
\begin{abstract}
I point out that the $U(N)$ Chern-Simons $3d$ theory coupled to fermions at finite temperature and at a specific mean field approximation and the $3d$ Gross-Neveu model at finite temperature and imaginary chemical potential can give us the same results for the thermodynamic values of the free-energy and the saddle point equation for the thermal mass. I further argue that the periodic structure of the imaginary chemical potential brings also Bloch's theorem into the game. Namely,  the vacuum structure of the fermionic system with imaginary baryon density is a Bloch wave. I further emphasise that Bloch waves correspond to fermionic (antisymmetric) or bosonic (symmetric) quasi-particles depending on the point in the band one sits in.

% Keywords - if any
\end{abstract}

\section{Introduction}
In recent years, the study of topological field theories has gained significant attention due to their intriguing mathematical structures and potential applications in condensed matter physics and high-energy physics (see e.g. \cite{Karch:2016sxi}-\cite{Meng:2020}). Among these theories, Chern-Simons gauge theory coupled to fermions (see for example \cite{Giombi:2011kc,Aharony:2012ns} and references therein) has emerged as a fascinating framework for investigating exotic phenomena and novel phases of matter \cite{Alvarez}. In this paper, I explore the interplay between Chern-Simons gauge theory and fermions at finite temperature, with a particular focus on the relationship with a fermionic theory at imaginary chemical potential.

I begin by providing a brief overview of Chern-Simons gauge theory and its relevance in various physical systems. I discuss the fundamental aspects of fermions coupled to Chern-Simons gauge fields, emphasizing their role in generating topological terms and inducing fractional statistics. Next, I introduce the concept of a finite-temperature formalism for the coupled system and present the necessary tools to describe the thermodynamics and transport properties of the theory and at the end I give new view into the partition function of a fermionic theory at imaginary chemical potential as a Bloch wave.

Overall, this paper provides a comprehensive investigation into the interplay between Chern-Simons gauge theory, fermions, and imaginary chemical potential at finite temperature. My results shed light on the rich physics underlying this coupled system and offer insights into potential applications in various areas of physics, including topological insulators, fractional quantum Hall systems, and high-energy physics.

\section{General notation}
\subsection{Maxwell}
To fix notation consider electromagnatism. Maxwell's equations are
\begin{equation}
\label{Maxw1}
\vec{\nabla}\vec{E}=\frac{\rho}{\epsilon_0}\,,\,\vec{\nabla}\times\vec{E}=-\frac{\partial\vec{B}}{\partial t}
\end{equation}

\begin{equation}
\label{Maxw2}
\vec{\nabla}\vec{B}=0\,,\,\vec{\nabla}\times\vec{B}=\mu_0\vec{j}+\epsilon_0\mu_0\frac{\partial\vec{E}}{\partial t}
\end{equation}
and from now on I set $\epsilon_0\mu_0=1/c^2$, $c=1$. To introduce relativistic notation I set 
\begin{equation}
\label{eta4}
\eta_{\mu\nu}=diag(-1,1,1,1)\,,\,\,\,\mu,\nu,\rho,..=0,1,2,3\,,\,\,\,i,j,k,..=1,2,3
\end{equation}
and define the field strength
\begin{equation}
\label{F}
F_{\mu\nu}=\partial_\mu A_\nu-\partial_\nu A_\mu
\end{equation}
The Maxwell action coupled to a source is
\begin{equation}
\label{Maction}
I=I_M+I_{int}=\int d^4x{\cal L}_{M} +\int d^4xA_\mu J^\mu=-\frac{1}{4}\int d^4 x F_{\mu\nu}F^{\mu\nu}+\int d^4x A_\mu J^\mu
\end{equation}
leading to the equation of motion
\begin{equation}
\label{Meom}
\partial_\mu F^{\mu\nu}=-J^\mu
\end{equation}
Define now
\begin{equation}
\label{EB}
\vec{E}\mapsto E^i=F^{0i}\,,\,\,\,\vec{B}\mapsto B^i=\frac{1}{2}\epsilon^{ijk}F_{jk}\Rightarrow F_{ij}=\epsilon_{ijk}B^k
\end{equation}
Notice that $\epsilon_{123}=\epsilon^{123}=1$. With the above notations I find
\begin{equation}
\label{EMaction}
I_M=\frac{1}{2}\int d^4x(\vec{E}^2-\vec{B}^2)
\end{equation}
and comparing with Maxwell's equations
\begin{equation}
\label{Jm}
J^\mu =\left(\frac{\rho}{\epsilon_0},\mu_0\vec{j}\right)
\end{equation}

\subsection{Abelian CS}
The Abelian CS action coupled to sources is
\begin{equation}
\label{CSaction}
I=I_{CS}+I_{int}=\int d^3x{\cal L}_{CS}+I_{int}=\frac{\kappa}{2}\int d^3x \epsilon^{\mu\nu\rho}A_\mu\partial_\nu A_\rho+\int d^3x A_\mu J^\mu
\end{equation}
Notice that now we are in three dimensions hence
\begin{equation}
\label{eta3}
\eta_{\mu\nu}=diag(-1,1,1)\,,\,\,\,\mu,\nu,\rho,..=0,1,2\,,\,\,\,i,j,k,..=1,2
\end{equation}
hence now $\epsilon_{012}=-\epsilon^{012}=1$. The first thing to notice is that the CS action does not depend on the metric. To see this we can write it as
\begin{equation}
\label{CSactionform}
\int_{{\cal M}}A\wedge dA=\int_{{\cal M}}(A_\mu\partial_\nu A_\rho )dx^\mu\wedge dx^\mu\wedge dx^\rho
\end{equation}
where ${\cal M}$ is a three dimensional Lorentzian manifold without a boundary for a start. Recall now that the invariant measure on ${\cal M}$ is
\begin{equation}
\label{dV}
dV_3=\epsilon^0\wedge e^1\wedge e^2=\sqrt{-g}dx^0\wedge dx^1\wedge dx^2=\sqrt{-g}d^3x
\end{equation}
Notice now that
\begin{equation}
\label{dx}
dx^\mu\wedge dx^\nu\wedge dx^\rho =\epsilon^{\mu\nu\rho}dx^0\wedge dx^1\wedge dx^2=E^{\mu\nu\rho}dV_3
\end{equation}
It is important to point out that $\epsilon_{\mu\nu\rho}$ takes values $\pm 1$ and it is {\it not} a tensor since
\begin{equation}
\label{e}
\epsilon^{\mu\nu\rho}=\frac{1}{(\sqrt{-g})^2}\epsilon_{\mu\nu\rho}
\end{equation}
The proper tensor density is $E_{\mu\nu\rho}$ defined as
\begin{equation}
\label{E}
E^{\mu\nu\rho}=\frac{1}{\sqrt{-g}}\epsilon^{\mu\nu\rho}
\end{equation}
From the above we see that the metric does not appear in the CS action, which is therefore topological.

Back to the CS action, the equations of motion are
\begin{equation}
\label{aCSeom}
\frac{\kappa}{2}\epsilon^{\mu\nu\rho}\partial_\nu A_\rho=-J^\mu\,,\,\,\,\frac{\kappa}{2}\epsilon^{\mu\nu\rho}F_{\nu\rho}=-J^\mu
\end{equation}
With the preview definitions these read
\begin{equation}
\label{aCSeom1}
B=\frac{\rho}{\kappa}\,,\,\,\,\epsilon^{ij}E_j=\frac{1}{\kappa}J^i\,,\,\,\,J^\mu=(\rho,J^i)
\end{equation}
Now, the magnetic field $B$ is a pseudoscalar.

\subsection{Non-Abelian Chern-Simons}

The gauge potential now transforms in the adjoint representation of the complex Lie algebra $SU(N)$ 
\begin{equation}
\label{A}
A_\mu=A^a T^a\,,\,\,\,[T^a,T^b]=f^{abc}T^c \,,\,\,\,Tr(T^a T^b)=\frac{1}{2}\delta^{ab}\,,\,\,\,a,b,c,...=1,2,...N^2-1
\end{equation}
The prime example is $SU(2)$ where $T^a=\frac{1}{2}\sigma^a$ and the structure constants $f^{abc}$ are complex. The non-abelian action is
\begin{equation}
\label{naCSaction}
{\cal I}_{CS}=\frac{\kappa}{4\pi}\int _{\cal M} Tr\left(A\wedge dA+\frac{2}{3}A\wedge A\wedge A\right)\nonumber \\
 =\frac{\kappa}{4\pi}\int_{\cal M}d^3x\epsilon^{\mu\nu\rho}Tr\left(A_\mu\partial_\nu A_\rho+\frac{2}{3}A_\mu A_\nu A_\rho\right)
 \end{equation}
 Since under the gauge transformation ($g$ is a group element of $SU(N)$)
 \begin{equation}
 \label{gtransf}
 A\mapsto g^{-1} Ag+g^{-1} dg
 \end{equation}
 the CS action is not invariant but changes as
 \begin{equation}
 \label{CSgtransf}
 {\cal I}_{CS}\mapsto {\cal I}_{CS}-2\pi\kappa \int_{\cal M} w(g)
 \end{equation}
 where $w(g)$ is the winding number
 \begin{equation}
 \label{wg}
 w(g)=\frac{1}{24\pi^2}\epsilon^{\mu\nu\rho} Tr\left(g^{-1}\partial_\mu g g^{-1}\partial_\nu g g^{-1}\partial_\rho g\right)
 \end{equation}
 then, invariance of the partition function
 \begin{equation}
 \label{pf}
 {\cal Z}=\int ({\cal D}A_\mu)e^{i{\cal I}_{CS}}
 \end{equation}
 requires that $\kappa$ is an integer $\kappa =N$, $N\in \mathbb{Z}$.

\section{The fractional quantum Hall effect in connection with Chern-Simons gauge field coupled to matter}

A Chern-Simons Lagrangian when matter is coupled with the gauge field is 
\begin{equation}
L_{CS}=\frac{\kappa}{2}\epsilon^{\mu\nu\rho}A_\mu\partial_\nu A_\rho+A_\mu J^\mu
\end{equation}
The Euler-Lagrange equations are (\ref{aCSeom})
where $J^\mu$ is the matter current $\left(\rho,J^i\right) $ and $i=1,2$ in $3$ dimensions.
The first component of the Euler-Lagrange equations is
\begin{equation}
\label{Flux}
\rho=\kappa B
\end{equation}
This equation tells us that the charge density is locally proportional to the magnetic field – thus the effect of a Chern-Simons field is to tie magnetic flux to electric charge $\left( anyon\right)$. So imagine a number N of particles tied with magnetic flux $\Phi$. The main point of this picture is that each particle sees each of the $N-1$ others as a point vortex of flux $\Phi$. This model is appropriate to describe the fractional quantum hall effect. This is a physical phenomenon in which the Hall conductance of 2D electrons shows precisely quantised plateaus at fractional values of $e^{2}/h$.
This statement becomes clearer from the calculation of the overall phase from the interaction between the charge $e$ of an anyon and the total flux of the other $anyons$ that travel adiabatically around it. Suppose that we have $N$ particles around this $anyon$ so since all particles travel around the $anyon$ the phase of the $anyon$ changes by $2\pi N$  . Then if we name the Berry phase of this exchange as $\gamma$ we will have
\begin{equation}
\gamma=2\pi N
\end{equation}
But for these $N$ particles we consider that the flux attached to each one of the particle is $\Phi$ and the flux quantum is $\Phi_0=\frac{h}{e}$ then
\begin{equation}
\gamma=2\pi N\frac{\Phi}{\Phi_0}
\end{equation}
It is interesting to remember the work \cite{Aharony:2012ns} where the calculations took into account that the eigenvalues of the Chern-Simons gauge field were living on the thermal circle having periodic characteristics like some electrons encircling an $anyon$. This $anyon$-eigenvalue was the result of the flux attachment that I mentioned before. So, since the eigenvalues of the Chern-Simons field behaved like electrons (the effective action of the model describes the eigenvalue like a particle in a magnetic field) they were quantized in units of $\frac{e}{\kappa}$, $\kappa$ is the Chern-Simons level and fill $\nu$ hypothetic Landau levels. The overall $\gamma$ phase is now
\begin{equation}
\gamma=2\pi N\frac{\frac{e}{\kappa}}{\Phi_0}
\end{equation}
where again $\Phi_0=\frac{h}{e}$. Having in mind that $\hbar=\frac{h}{2\pi}$ then the final value of the overall phase turns to
\begin{equation}
\gamma=N\frac{e^2}{\hbar\kappa}
\end{equation}
At \cite{Dunne} for $\hbar=1$ the $\gamma$ phase is now
\begin{equation}
\gamma=N\frac{e^2}{\kappa}
\end{equation}
So when an electron moves adiabatically around another electron its wavefunction acquires a phase like:
\begin{equation}
\psi'=e^{iN\frac{e^2}{\kappa}}\psi
\end{equation}
or by using the 't
Hooft coupling $\lambda =\frac{N}{\kappa}$ we have:
\begin{equation}
\psi'=e^{iN\frac{\lambda e^2}{N}}\psi
\end{equation}
and for simplicity if we set $e^2=1$ we have finally:

\begin{equation}
\label{Bohm}
\psi'=e^{i\lambda}\psi
\end{equation}
This picture corresponds to electrons encircling a thin solenoidal magnetic flux and so we expect to encounter Bohm-Aharonov type phenomena (\cite{Christiansen:1999uv},\cite{Huang}). A similar picture is to have fermions at imaginary chemical potential as we will examine later.

\section{Imaginary chemical potential vs Chern-Simons gauge field}
Returning to equation (\ref{Flux}) one may think the possibility to use an electric field instead of a magnetic one to achieve an electric flux attachment this time. Having in mind that in $2+1$ dimensions physics coming from the coupling of a Chern-Simons gauge theory with fermionic matter, the two components of the electric field lie on the same plane in contrast with the one component of the magnetic field which is perpendicular to the plane, we may suppose that from a fermionic model at a real chemical potential that lies on the same plane as the Chern-Simons electric components one may rotates it until it comes on the imaginary axis. Then the model turns to a fermionic model with imaginary chemical potential where we can study statistical transmutation but also some amazing similarities with the above models. Since physics with an electromagnetic field like the Chern-Simons field is much more complicated than physics with an imaginary electric potential our main ambition is to examine the condition:

 "Gross Neveu model at critical values of the imaginary chemical potential$\rightarrow$ Fractional quantum Hall effect at critical values of the $\nu$ Landau levels". 
 
 Of course a model with a chemical potential is closed to a model with an electric field in a crystal and polarization phenomena from the Berry phase coming from the periodic structure of the crystal. So if one imagines the movement of an electron in a crystal (along for example the first Brillouin zone) a neighbor electron somehow travels around it because a Brillouin zone in one dimension can be mapped onto a circle, in view of the fact that wavevectors $k=0$ and $k=\frac{2\pi}{a}$ label the same states. 

An equivalent picture is the above:
When fermions are coupled to a Chern-Simons gauge field in a monopole background, their system exhibits intriguing properties such as the emergence of anyonic statistics, which refers to statistics that are neither fermionic nor bosonic but can be fractional. The study of theories like this finds applications in condensed matter physics, like the FQHE and topological insulators.

At finite temperature, (see e.g. \cite{ZinnJustin:2002ru}) and 
in \cite{Filothodoros:2016txa}-\cite{Thesis Filothodoros} it was argued that for Dirac fermions coupled to an abelian Chern-Simons field at level $\kappa$ in three Euclidean  dimensions\footnote{My notations follow \cite{ZinnJustin:2002ru}.} the presence of a monopole charge is closely related to the presence of an imaginary chemical potential. Let's see the connection:
\begin{align}
\label{DiracCSPF}
Z_{fer}(\beta,\kappa)&=\int [{\cal D}A_\nu][{\cal D}\bar\psi][{\cal D}\psi]\exp{\left[-S_{fer}(\bar\psi,\psi,A_\nu)\right]}\,,\\
\label{Sf}
S_{fer}(\bar\psi,\psi,A_\nu)&=-\int_0^{\beta}\!\!\!d\tau\!\!\int \!\!d^2\bar{x}\left[\bar{\psi}(\slash\!\!\!\partial -i\slash\!\!\!\!A)\psi+i\frac{\kappa}{4\pi}\epsilon_{\nu\lambda\rho}A_\nu\partial_\lambda A_\rho+...\right]\,.
\end{align}
There are also some fermionic self interactions that their presence presented with dots. When one expands the CS field around time independent monopole configuration $\bar{A}_\nu$ \cite{Fosco:1998cq}
\begin{equation}
\label{Aexpansion}
A_\nu=\bar{A}_\nu+b_\nu\,,\,\,\,\bar{A}_\nu=(0,\bar{A}_1(\bar{x}),\bar{A}_2(\bar{x}))\,,\,\,\, b_\nu=(b_0(\tau),b_1(\tau,\bar{x}),b_2(\tau,\bar{x}))\,,
\end{equation}
which is normalized as\footnote{For example, one may consider the theory on $S^1\times S^2$.}

\begin{equation}
\label{monopole}
\frac{1}{2\pi}\int d^2x \bar{F}_{12}=1\,,\,\,\,\bar{F}_{\nu\lambda}=\partial_\nu \bar{A}_\lambda-\partial_\lambda \bar{A}_\nu\,
\end{equation}
and $b_\nu$ is a backround gauge field.
Hence, (\ref{Sf}) describes the possibility of monopole configurations in the fermionic theory (or the attachment of $\kappa$ units of monopole charge to the fermions) as
\begin{equation}
\label{Sfexp}
S_{fer}(\bar\psi,\psi,A_\nu) =-\int_0^\beta \!\!\!d\tau\!\!\int \!\!d^2\bar{x}\left[\bar\psi(\slash\!\!\!\partial-i\gamma_i\bar{A}_i-i\gamma_\nu b_\nu)\psi+i\frac{\kappa}{4\pi}\epsilon_{\nu\lambda\rho}b_\nu\partial_\lambda b_{\rho}+..\right]-i\kappa\int_0^\beta \!\!\!d\tau b_0\,.
\end{equation}
We can perform the path integral over the CS fluctuations like viewing the theory with fixed total monopole charge. To do this, I use a mean field approximation in this sector where the spatial CS fluctuations balance out the magnetic background gauge filed  $\langle b_i\rangle =-\bar{A}_i$ \cite{Barkeshli:2014ida}.\footnote{It would be interesting to further explain whether an appropriate large-$N$ is also necessary for the validity of such an approximation.} It is comperable to how we may think of a reduction in which the background gauge field's  integral along the thermal circle is fixed to be a constant. I then obtain
\begin{align}
Z_{fer}(\beta,\kappa)&=\int [{\cal D} b_0][{\cal D}\bar\psi][{\cal D}\psi]\exp{\left[\int_0^\beta \!\!\!d\tau \!\!\int \!\!d^2\bar{x}\left[\bar\psi(\slash\!\!\!\partial-i\gamma_0 b_0)\psi+..\right]+i\kappa\int_0^\beta \!\!\!dx^0 b_0\right]}\nonumber \\
\label{DiracCSPFfin}
&=\int ({\cal D}\theta)e^{i\kappa\theta}Z_{gc,fer}(\beta,-i\theta/\beta)\,,
\end{align}
where  $\theta=\int_0^\beta d\tau b_0(\tau)$, $Z_{gc,fer}(\beta,-i\theta/\beta)$ is the grand canonical partition function for the fermionic theory and I have used standard formulae from \cite{ZinnJustin:2002ru}. We see that the CS level $\kappa$ plays the role of the eigenvalue $Q$ of the $U(1)$ charge operator. 
Here is an important conclusion that the finite temperature partition function of Dirac fermions coupled to abelian CS gauge field at level $\kappa$ in a monopole background, is equivalent to the canonical partition function of the fermions at fixed fermion number $\kappa$.
This discussion shows that the partition function of fermions coupled to abelian CS in a monopole background is intimately related to the respective canonical partition function at fixed total $U(1)$ charge.

Let's compare the two theories by using previous results.

Here we have a theory of N Dirac fermions $\psi ^{i}$ $(i=1,....,N)$ coupled
to $U(N)$ Chern-Simons gauge field $A_{\mu }$. The theory is defined by the
Lagrangian density

\begin{equation}
L_{CS(f)}=\overline{\psi} ^{i}D_{\mu }\psi ^{i}+\sigma \overline{\psi} ^{i}%
\psi ^{i}+\frac{i\kappa}{4\pi }\varepsilon ^{\mu \nu \rho }Tr(A_{\mu }\partial
_{\nu }A_{\rho }+\frac{2}{3}A_{\mu }A_{\nu }A_{\rho })
\end{equation}

where $D_{\mu }\psi =\partial_{\mu }\psi -iA^{a}T^{a}\psi.$

When the theory is critical (The Gross-Neveu model with Chern-Simons
interactions at the saddle point) $\sigma $ plays the role of the primary
scalar operator and we must perform the path integral over it. Previous work \cite{Aharony:2012ns}
was made by using the light-cone gauge to eliminate the non-abelian part of
the Lagrangian. Since in the thermal theory the fermions have anti-periodic
boundary conditions on the thermal circle, the momenta in this direction are
quantized as $p_{3}=\frac{2\pi }{\beta}(n+\frac{1}{2}),n\in\mathbb{Z}.$

The free-energy density of the model is (\cite{Aharony:2012ns}):

\begin{equation}
\frac{F_{CS}}{V_{2}}=\frac{N}{2\pi \beta^{3}}\left\{ \frac{\mu _{F}^{3}}{3}\left(
1\mp \frac{1}{\lambda }\right) +\frac{\widetilde{\sigma }\mu _{F}^{2}}{%
2\lambda }-\frac{\widetilde{\sigma }^{3}}{6\lambda }+\frac{1}{\pi i\lambda }%
\int_{\mu _{F}}^{\infty }dyy\left[ Li_{2}\left( -e^{-y+\pi i\lambda }\right)
-c.c.\right] \right\} 
\end{equation}

where $\widetilde{\sigma}=\sigma \beta$ and $\lambda =\frac{N}{\kappa}$ is the 't
Hooft coupling.

When the theory is critical we extremize the $\frac{F_{CS}}{V_{2}}$ with
respect to $\widetilde{\sigma }$ and we find

\begin{equation}
\widetilde{\sigma }=\pm \mu _{F} 
\end{equation}

With the appropriate choice of signs (the (-) sign in the parenthesis and
the (+) sign at the above equation) the critical free-energy density we
obtain is

\begin{equation}
\frac{F_{CS}}{V_{2}}=\frac{N}{2\pi \beta^{3}}\left\{ \frac{\mu _{F}^{3}}{3}+\frac{%
1}{\pi i\lambda }\int_{\mu _{F}}^{\infty }dyy\left[ Li_{2}\left( -e^{-y+\pi
i\lambda }\right) -c.c.\right] \right\} 
\end{equation}

and the saddle point equation is

\begin{equation}
\lambda \mu _{F}=-\frac{1}{\pi i}\left[ Li_{2}\left( -e^{-\mu _{F}-\pi
i\lambda }\right) -c.c.\right] 
\end{equation}

The main difference here with the $2+1$ Gross-Neveu model in the canonical formalism from previous calculations is to include the
integral over all the possible $A_{0}$ values ($A_{0}=\frac{2\pi nc}{e}$
where $c\epsilon \left[-1/2,1/2\right]$). The corresponding results are (\cite{Filothodoros:2016txa}):

\begin{equation}
F_{GN}=\frac{NV_{2}}{2\pi }\left\{ \frac{m^{3}}{3}+\int_{0}^{\infty
}pdp\left( \frac{1}{i\pi \left( \frac{n}{e}\right) }\right) \left[
Li_{2}\left( -e^{-L\sqrt{p^{2}+m^{2}}+i2\pi \left( \frac{n}{e}\right)
}\right) -Li_{2}\left( -e^{-L\sqrt{p^{2}+m^{2}}-i2\pi \left( \frac{n}{e}%
\right) }\right) \right] \right\} 
\end{equation}

Obviously the expressions of the free-energies of the two models are the
same if we consider that
\begin{equation}
\mu _{F}=m\beta
\end{equation}
and
\begin{equation}
\lambda =\frac{n}{e}
\end{equation}

and also
\begin{equation}
m=-\frac{1}{\pi i\left( \frac{n}{e}\right) }\left[
Li_{2}\left( -e^{-m\beta-\pi i\left( \frac{n}{e}\right) }\right) -c.c.
\right] 
\end{equation}

In principle, the free-energy $F_{GN}$ together with the gap equation are
sufficient for studing the thermodynamic properties of the Gross-Neveu model
to leading $N$. 

 However the main conclusion from the
above analysis is that as the Chern-Simons gauge field "ties" charge with
flux by the Chern-Simons $\kappa$ level and somehow we have fixed number of
particles, the constraint that we insert at the Gross-Neveu model tells us
that we have also fixed number of particles. One may assume this from the
beginning of our analysis by examine carefully the Lagrangians of the models
and find the equivalence of the parts
\begin{equation}
L_1=i\kappa b_0
\end{equation}

which is a (flux)$\times $(gauge potential as magnetic field) and
\begin{equation}
L_2=iB\theta
\end{equation}

which is also a (electric flux arises from the charge of particles)$\times $(potential) \cite{Filothodoros:2016txa}. So we may say that with the constraint we put in the Lagrangian
we "create" something like a topological gauge field.

\section{Partition function and Bloch waves}

The grand canonical partition function of a Dirac fermion\footnote{For simplicity in this section I consider theories with a single fermion or complex scalar.} in three dimensions and in the presence of an imaginary chemical potential can be written as
\begin{equation}
\label{1Dirac}
Z^{f}_{gc}(\beta,i\theta /\beta)=\int ({\cal D}\psi)({\cal D}\bar\psi)e^{-\int_0^\beta d\tau\int d^2\vec{x}\left[\bar\psi(\gamma^\mu\partial_\mu -i\gamma^0\theta/\beta)\psi+V(\bar\psi\psi)\right]}
\end{equation}
where momentarily we are not interested in the potential term $V(\bar\psi\psi)$. Placing the fermions at finite temperature requires imposing antiperiodic boundary conditions along the compactified Euclidean time $\tau$ as
\begin{equation}
\label{apBC}
\psi(\beta,\bar{x})=-\psi(0,\bar{x})\,,\,\,\,\bar{\psi}(\beta,\bar{x})=-\bar{\psi}(0,\bar{x})
\end{equation}
The presence of the imaginary chemical potential in (\ref{1Dirac}) is actually equivalent to the coupling of the fermions to a background gauge potential of the form $A_{\mu}=(\theta/\beta,0,0)$. One therefore might think that this is removable by a simple gauge transformation, and setting
\begin{align}
\psi(\tau,\bar{x})\mapsto \psi'(\tau,\bar{x})=e^{i\int_0^{\tau} d\tilde{\tau}\alpha_0(\tilde{\tau})}\psi(\tau,\bar{x})\,,\,\,\\\bar\psi(\tau,\bar{x})\mapsto \bar\psi'(\tau,\bar{x})=e^{-i\int_0^{\tau} d\tilde{\tau}\alpha_0(\tilde{\tau})}\bar\psi(\tau,\bar{x})\,.
\end{align}
could do the job. However, such a transformation would twist the antiperiodic boundary conditions (\ref{apBC}) since now we would obtain
\begin{equation}
\label{twistapBC}
\psi(\beta,\bar{x})=-e^{-i\theta}\psi(0,\bar{x})\,,\,\,\,\bar{\psi}(\beta,\bar{x})=-e^{i\theta}\bar{\psi}(0,\bar{x})
\end{equation}
where these twists are intimately related to the confining/deconfining of colour singlets. Obviously there is a connection with (\ref{Bohm}) and this is a main result of this work. Also, if one consider $\psi$ as a Bloch wave then $\theta$ could change the "lattice" periodic part of the wavefunction.

Then I discuss the theory for imaginary chemical potential. The important point is that this situation corresponds to having the GN model coupled to a $U(1)$ potential fluctuating around its $\mathbb{Z}$-vacua. This is the proposal of my previous work \cite{Filothodoros:2016txa}.
The fact that I use an imaginary chemical potential brings the Lee-Yang theorem into the game. We need to elaborate on that, in particular since I find {\it negative} free-energy densities. This shows that we probe unstable critical points. But this is not the main case in this work.

The situation resembles studies of quantum mechanical systems in a periodic potential like a periodic crystal. If we think of $\theta$ as a periodic coordinate, then is equivalent to the calculation of the overlap between two Bloch wavefunctions that differ by  lattice momentum $B$ \cite {Blochoverlaps}. Such systems usually generate a band structure which can be studied by restricting the lattice momentum to the first Brillouin zone. 

Consider a system in three dimensions and at finite temperature $T=1/\beta$ with a global $U(1)$ charge operator $\hat{Q}$. Its canonical partition function can be formally calculated as the thermal average over states with fixed $\hat{Q}$ as
\begin{equation}
\label{canPF}
Z_c(\beta,B)=Tr\left[\delta(\hat{Q}-B)e^{-\beta\hat{H}}\right]
\end{equation}
If the eigenvalues $B$ of $\hat{Q}$ are integers, namely if the system contains elementary excitations, an explicit  representation of (\ref{canPF}) can be written as
\begin{equation}
\label{gcPF}
Z_c(\beta,B)=\int_{0}^{2\pi}\!\frac{d\theta}{2\pi}\,e^{i\theta B}\,\rm Tr\left[e^{-\beta\hat{H}-i\theta\hat{Q}}\right]=\int_{0}^{2\pi}\!\frac{d\theta}{2\pi}\,e^{i\theta B}\,Z_{gc}(\beta,i\mu=i\theta/\beta),
\end{equation}
where $Z_{gc}(T,i\mu)$ is the grand canonical partition function with imaginary chemical potential $i\mu$. 

  In the simple systems we are interested in one expects that 
\begin{equation}
\label{Zgcperiodic}
Z_{gc}(\beta,i(\mu +2\pi k /\beta))= Z_{gc}(\beta,i\mu)\,,\,\,\,k\in \mathbb{Z}
\end{equation}
One then notices with Bloch's theorem as follows. In quantum system, taken here to be $1d$ for clarity,  in a periodic potential with period $a$ the energy eigenstates are the Bloch waves
\begin{equation}
\label{Blochwaves}
\psi_k(x) =e^{ikx}u(x)\,,\,\,\,u(x+a)=u(x)
\end{equation}
where $k$ is the lattice momentum vector. The transition amplitude between two Bloch waves with different lattice momenta is
\begin{equation}
\label{transition}
Z(k_2-k_1)\equiv \langle \psi_{k_1}|\psi_{k_2}\rangle =\int_0^a dx e^{i(k_2-k_1)x}|u(x)|^2
\end{equation}
Notice the formal equivalence of (\ref{transition}) with (\ref{gcPF}), which implies that $B$ may be thought of as a transfer momentum when a Bloch wave scatters from a lattice point \cite{Filothodoros:Bloch}. The appropriate expression of a $1d$ Bloch-wave is of the form:

\begin{equation}
\psi_{k}=e^{ikx} u(x)
\end{equation}
where $u$ function has the periodicity of the lattice $a$ like $Z_{gc}(\beta,i\mu=i\theta/\beta)$ has the periodicity of the chemical potential.

\section{Discussion}

I have pointed out that the canonical partition function of the Gross-Neveu model at finite temperature and imaginary chemical potential is intimately related to the thermal partition function of abelian Chern-Simons fields coupled to matter in a monopole background, when a suitable mean-field approximation is assumed. One may think of the latter as the regime where the $U(1)$ charge density essentially corresponds to the monopole charge. Motivated from my previous work \cite{Filothodoros:Bloch} about the relationship between the partition function of the fermionic model and the Bloch theorem I shed more light to the connection between the periodic structure of the imaginary chemical potential on the thermal circle and a periodic band structure where a Bloch wave travels. It would also be interesting in the future to examine the specific values of the imaginary chemical potential where there are phase transitions of the model and their relationship with the fractional quantum Hall effect. 

\section*{Acknowledgements}
I would like to thank A. C. Petkou for his useful comments and help.

\end{document}